# Ultrasound-Triggered Release of Anticancer Nanoparticles from Electrospun Fabrics Integrated with Soft Robotic Tentacles

*Samuel C. T. Moorcroft,\* Benjamin Calmé, Charles Brooker, Pietro Valdastri, Russell Harris, Stephen J. Russell, Giuseppe Tronci*

S. C. T. Moorcroft, C. Brooker, S. J. Russell, G. Tronci
Clothworkers' Centre for Textile Materials Innovation for Healthcare, Leeds Institute of Textiles and Colour, School of Design, University of Leeds, Leeds, U.K.
E-mail: texsmo@leeds.ac.uk

B. Calmé, P. Valdastri
Science and Technology of Robots in Medicine (STORM) Laboratory, School of Electronic and Electrical Engineering, University of Leeds, Leeds, U.K.

R. A. Harris
Department of Engineering, Manchester Metropolitan University, Manchester, U.K.

C. Brooker, G. Tronci
Division of Oral Biology, School of Dentistry, St. James's University Hospital, University of Leeds, Leeds, U.K.

The prompt identification of pancreatic cancer symptoms is an ongoing clinical challenge, often leading to late diagnosis and poor prognosis. Tumor 'hijacking' of the pancreatic stromal structure limits the uptake of systemic chemotherapeutics. Localized drug delivery systems (DDS) using endoluminal techniques are widely utilized, with positive early results for improved control over tumor growth. There is a need for technologies that integrate endoluminal resources and intelligent material systems to better control the spatiotemporal delivery of chemotherapeutics. We demonstrate the ultrasound (US)-triggered localized release of therapeutics through the design of solvent traceless drug-loaded vinylbenzyl-functionalized gelatin (gel4vbc) nanoparticles (NPs) integrated with an electrospun fabric. Albumin-loaded

NPs encapsulated into a poly(vinyl alcohol) (PVA) coating of a poly($\varepsilon$-caprolactone) fabric evidence tunable triggered NP delivery controlled by regulating PVA concentration (0–1 wt.%) and US intensity (0–3 W·cm$^{-2}$). The fixation of the NP-coated fabric to a magnetic tentacle robot (MTR) enables the automated manipulation into a phantom pancreatic duct before the US-triggered release of NP-loaded albumin and MTR retraction. Albumin release is controlled by varying the surface area of the NP-loaded MTR-coating fabric. Herein we have designed a novel DDS capable of facile integration into soft robotics and US-triggered delivery of therapeutic-loaded NPs.

**Key Points**

- Fabrication of photocurable gelatin nanoparticles using a nanoemulsion technique with traceless solvent post fabrication.
- Incorporation of therapeutic-loaded nanoparticles into an electrospun fabric and subsequent fixation to a magnetic tentacle robot.
- Automated manipulation of nanoparticle-loaded fabric coated magnetic tentacle robot into a phantom pancreatic duct, followed by controlled and tunable ultrasound-triggered release of therapeutics, and tentacle retraction.

**Graphical abstract**

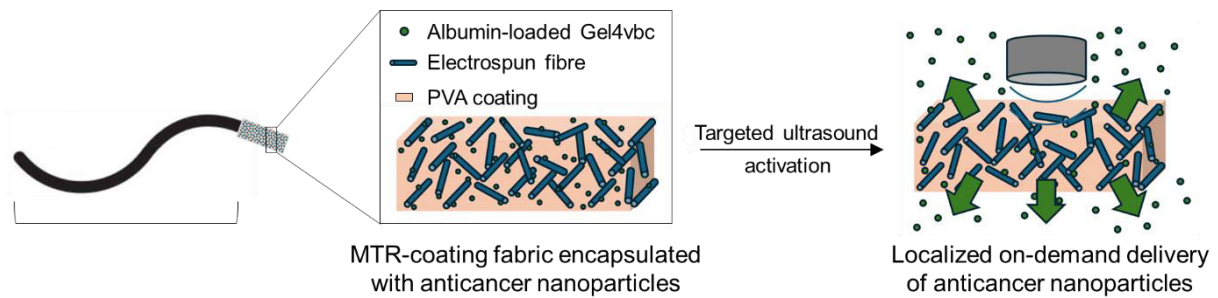

## 1. Introduction

Delayed diagnosis and resistance to chemotherapeutics make pancreatic cancer one of the deadliest and most challenging cancers to combat, with treated patients having a median survival duration of 10 – 12 months.[1–3] Vague symptoms cause late detection at an advanced stage, meaning the microenvironment has already been altered by the cancer, creating a favorable environment for tumor growth.[4] The surrounding stroma tissue is 'hijacked', accounting for over 70 % of the tumor volume, leading to vascular compression and restricted accessibility of chemotherapeutics and immune system cells.[5,6] As such, chemotherapeutic delivery through the bloodstream is inhibited, making systemic drug administration largely ineffective. Consequently, localized delivery of active agents is growing increasingly popular.[7–10]

Site-specific delivery minimizes side effect toxicity risks and enables increased drug retention at the tumor site[11] while using a lower dosage reduces the toxic stresses imparted by anti-cancer chemotherapeutics.[12] Early evidence of intertumoral and peritumoral delivery of chemotherapeutics using fine needle injection[13] and endoscopic ultrasound[14,15] indicate potential for control over tumor growth and enhanced quality of life. Such localized delivery techniques offer increased long-term survival compared to open surgery,[16,17] illustrating a need for technology platforms integrating novel, patient-friendly techniques and stimulus-responsive drug delivery systems (DDSs) with enhanced spatiotemporal control.

Magnetic tentacle robots (MTRs) are thin, soft, scalable, and can be shaped into curvilinear morphologies under magnetic influence to enable navigation within the human cavities. As such, MTRs are excellent candidates for carrying out endoluminal drug delivery,[18–20] whereby their magnetic responsivity can be leveraged to carry the DDS directly to a tumor site, allowing for localized agent release through activation of external stimuli, and to enable MTR retraction to minimize risks of foreign body response.[21] Stimuli-responsive delivery, therefore, plays a major role in this anticancer therapeutic modality, whereby the retention of the bulk of the therapeutic within the DDS-equipped MTR should be ensured until the targeted tumor region is reached and prior to the stimulus-initiated payload release. Exogenous stimuli (US, magnetic field, temperature, and photo-irradiation[22]) can induce precise spatiotemporal release,[23–25] enabling burst release of encapsulated molecules on-demand and therefore increasing the likelihood of delivery of therapeutic doses. Localized delivery to pancreatic tumors is already displaying promising results *in vitro* and *in vivo*; however, further investigation is required to demonstrate efficacy in clinical trials.[26–29]

Recent advances in material science and nanotechnology have facilitated the development of various intelligent DDSs customizable to specific clinical applications.[30–34] DDS utilizing gelatin polymeric scaffolds are widely adopted in healthcare research,[35–38] primarily due to gelatin's amino acidic composition, biodegradation in biological environments, low cost, and minimal risks of antigenicity.[39–41] The linear and amphoteric structure of gelatin enables chemical accessibility and facile functionalization aiming to control the release of internalized molecules and ensure material integrity in the presence of external stimuli.[42,43] Despite these advantages, the development and clinical translation of gelatin-based DDSs has been limited, partially due to the toxicity of crosslinking agents and poor drug retention *in vivo*, owing to uncontrolled swelling, rapid enzymatic degradation, and reversible gelation at physiological temperatures.[44,45] Fabrication of gelatin nanostructures often requires substantial quantities of stabilizing agents and toxic solvents, both of which can require complex post-processing removal.[46]

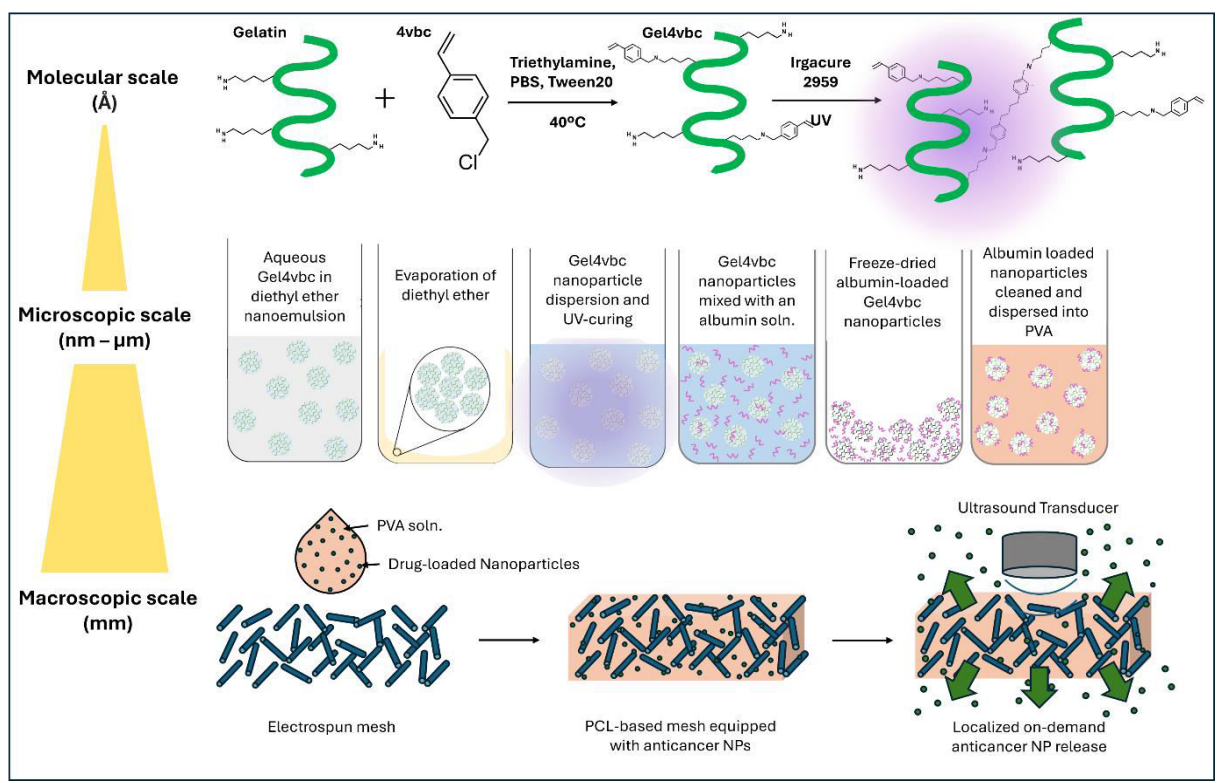

**Scheme 1.** Multiscale design of the US-triggered fabric for localized on-demand anticancer release. Molecular scale: gelatin is covalently functionalized with 4vbc to generate a UV-curable particle-forming system. Microscopic scale: 4vbc-functionalized gelatin (gel4vbc) is processed into UV-cured albumin-loaded NPs prior to dispersion onto a PVA solution. Macroscopic scale: A fibrous electrospun fabric is coated with the NP-loaded PVA solution to enable particle release on-demand under ultrasound (US).

We describe the multiscale design of drug-loaded NPs and their encapsulation into a polymer coated fabric (**Scheme 1**) for integration into a soft MTR system, providing a simple means to accomplish on-demand and localized drug delivery. We have previously shown the cellular tolerability and retained wet-state mechanical competence of electrospun fibers made of UV-cured gel4vbc.[47] By harnessing the scalability and manufacturability of this product, here we investigate the manufacture of gel4vbc NPs using a novel traceless solvent nano-emulsion process with simple post-process purification. The NPs are then coated onto a nanofibrous electrospun poly(ε-caprolactone) (PCL) fabric to create an US-sensitive device that can easily be integrated with a soft robotic tentacle. PCL electrospun fabrics, with constituent fiber diameters ranging from the nano- up to the microscale, are ideal scaffolds for *in vivo* applications due to the inherent safety and biodegradability, hydrophobicity, flexibility, and structural integrity at the macroscale.[48–52] By leveraging these features, we load the NP-loaded fabric onto a magnetic tentacle robot capable of manipulation through deep brain tissue,[53] peripheral nodules in the lungs,[54] and the complex architecture of the pancreas and gallbladder.[55] The DDS system is then guided through a duodenum phantom to demonstrate US-triggered albumin delivery within the pancreatic duct. Although albumin itself does not inherently possess anticancer properties, it is widely utilized as a carrier molecule for anticancer drugs, as it significantly enhances pharmacokinetic profiles, thereby improving both the efficacy and safety of therapeutic agents. Consequently, we selected albumin as a surrogate for anticancer drugs in this study.[56] US is selected as a safe, clinically compliant mechanical stimulus for drug release activation, aiming to independently control anticancer delivery and magnetic navigation of the MTR.

## 2. Results and Discussion
### 2.1. Gelatin functionalization with 4vbc and nanoparticle fabrication

The manufacture of UV-curable NPs consisting of gel4vbc is presented (Scheme 1). The degree of functionalization of gel4vbc is 40 ± 8 % (**Figure S1, Supporting Information**), in line with previous reports.[47] Increasing the molar excess of 4vbc and TEA yields gelatin products with a significantly increased degree of functionalization ($F$), so that an average value of $F$ of nearly 80% is reached with a 4vbc/Lys molar ratio of 100 (Figure S1). On the other hand, increasing the molar ratio of 4vbcLys in the range of 25-100 while keeping the TEA/Lys ratio constant at 25 does not lead to statistically significant differences in $F$, indicating the major role of TEA in enhancing the reactivity of the terminal primary amino groups of gelatin lysines. Despite the aforementioned changes in $F$, functionalization of the hydrophilic primary amino groups of lysine with hydrophobic 4vbc residues generates a gelatin product with decreased solubility at neutral pH, which is a hindrance for further processing. Consequently, gelatin products with an average $F$ of 40% are selected for further investigation.

Crosslinking of gelatin is vital for utilization in physiological conditions, due to the onset of the sol-gel transition of native gelatin hydrogels being 26°C, as ascertained with differential scanning calorimetry (**Figure S2**). The initial addition of the 4vbc moiety increases the sol-gel transition onset temperature to 37°C; however, post UV-curing of the gel in the presence of a photo-initiator (I2959), the onset of the glass transition is not identified beneath 100°C, which is evidence of UV-induced covalent crosslinking.[59,60] This finding corroborates with observations that cured gel4vbc gels maintain structural integrity when incubated at 100°C for over 15 mins (**Figure S3**).

NP fabrication is performed by dissolving gel4vbc into an acidic solution (10 mM HCl) and dropwise addition to a SPAN® 80–supplemented diethyl ether solution (Scheme 1). The emulsion is then homogenized for 2 min and mixed until the diethyl ether has evaporated away, leaving a paste of gel4vbc NPs. The NPs are then dispersed into a solution containing I2959 (1 wt%) and UV-cured for 30 min. The UV-induced crosslinking is critical to the formation of NPs, which is not possible using native gelatin, given that the particles swell in the aqueous environment to millimeter diameters.

The size of the NPs is primarily dictated by the shear forces applied during the homogenization step following formation of the emulsion (**Figure S4**). Without homogenization, mixing only with a magnetic stir bar at 1000 rpm generates micron-sized particles (Ø= 1022 ± 411 nm) along with millimeter-sized aggregates of coagulated polymer. On the other hand, homogenization at 25,000 rpm promotes the formation of smaller NPs (Ø=75 ± 44 nm). Aside from the homogenization rate, variations in homogenization duration do not significantly affect NP size (Figure S4), whereby an average diameter of approximately 200 nm is measured across a threefold increase in homogenization time (60–180 s).

The effects of the emulsion parameters, i.e. gel4vbc concentration, gel4vbc solution: diethyl ether volume ratio, and concentration of SPAN® 80, on particle size are also investigated (**Figure 1A**). Increasing the gel4vbc concentration in the aqueous phase from 10 wt% to 15 wt.% has a direct impact on the NP diameter, corresponding to a statistically significant size increase from 183 ± 61 nm to 414 ± 56 nm, respectively. No statistically significant change in size is observed between NPs containing 5 and 10 wt% gel4vbc, likely due to the lower viscosity enabling more consistent homogenization of the emulsion.[61] Altering the volume ratio between the aqueous and the solvent phase also has a negligible effect on the NP size, whereby all mean diameters are recorded in the range of 174 - 200 nm (Figure 1A). However, aqueous phase: solvent ratios greater than 1:16 yield coalescence of the aqueous phase and inhibition of NP formation. This observation therefore indicates that the selected range of aqueous-to-organic solvent volume ratios is optimal, aiming to accomplish stable NPs with controlled diameter. Furthermore, considering that the average lumen diameter of constricted capillaries in the exocrine pancreas—the smallest capillaries—is approximately 4 μm,[62] the gel4vbc NPs, being roughly 20 times smaller, should readily infiltrate the entire volume without obstruction.

The inclusion of a surfactant is also integral to the fabrication of NPs. Without SPAN® 80, the aqueous phase immediately and irreversibly coalesced, even with homogenization.

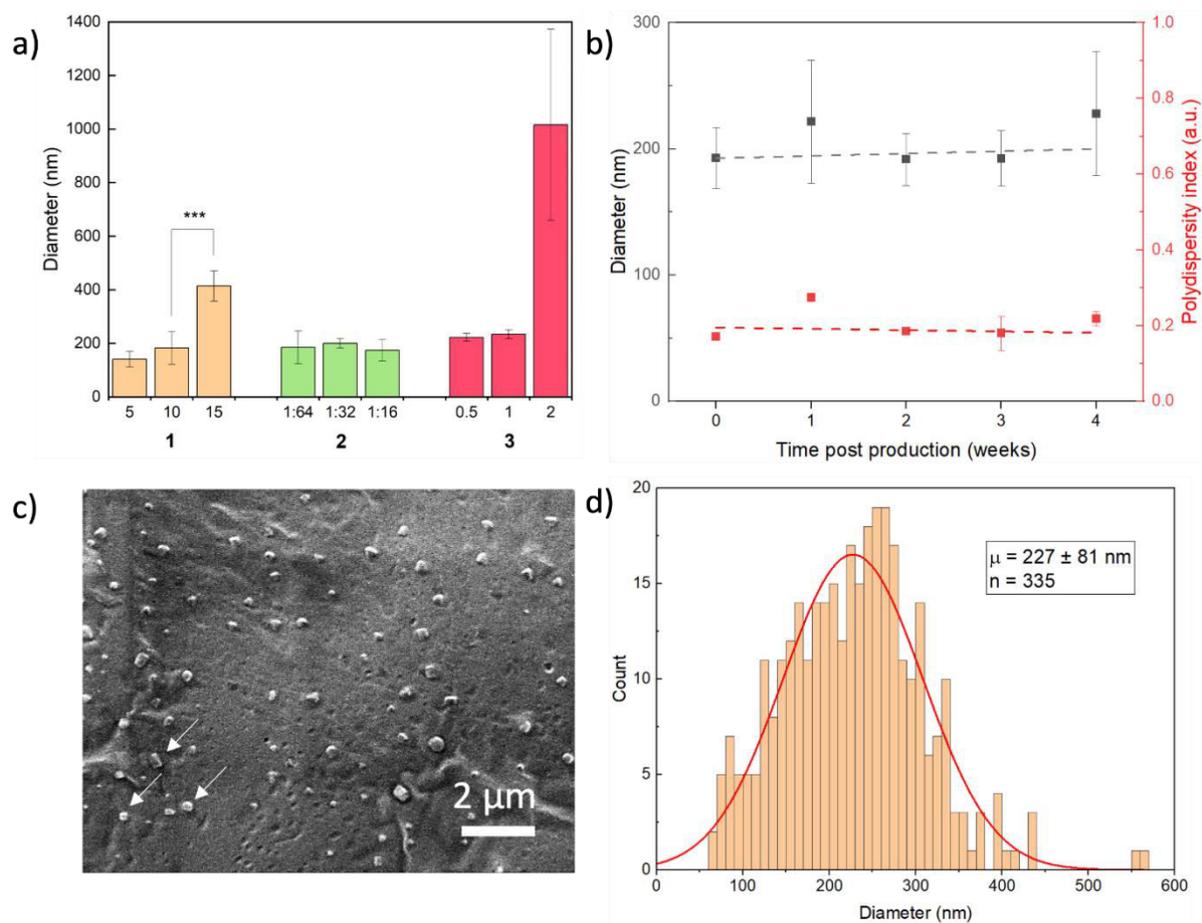

**Figure 1.** Control of particle size in varied emulsion and post-emulsion conditions. A): Diameters of gel4vbc NPs obtained under variation of: (**1**, orange) gel4vbc concentration (wt%) in the aqueous phase, with constant (1.32) aqueous phase : organic solvent volume ratio and constant (0.1 wt%) SPAN80 concentration, (**2**, green) aqueous phase : organic solvent phase volume ratio, using constant concentrations of gel4vbc (10 wt%) and SPAN80 (0.1 wt%), (**3**, red) SPAN80 concentration within the solvent phase, whereby 10 wt% of gel4vbc and 1:32 volume ratio of aqueous phase : organic solvent were constantly used. B): Diameter and polydispersity indices of gel4vbc NPs over a 4-week incubation period (10 mM HCl, 20-22°C, as determined using multi-angle dynamic light scattering. C): Scanning electron micrographs of freeze-dried gel4vbc NPs. White arrows indicate examples of the nanoparticles. D): Histogram of the diameters of freeze-dried gel4vbc NPs, determined through image analysis of scanning electron micrographs. In all figures, error bars indicate the standard deviation (n = 3). p-values indicate: *** ≤ 0.001.

The inclusion of a surfactant is also integral to the fabrication of NPs. Without SPAN® 80, the aqueous phase immediately and irreversibly coalesced, even with homogenization. On the other hand, relatively high concentration of surfactant in the organic solvent phase cause rapid NP aggregation. At 2 wt% SPAN® 80, NPs are found to aggregate into micrometer-sized structures during curing (Ø= 1026 ± 357 nm, Figure 1A), yielding covalent crosslinks between particle-forming gelatin chains and mechanical stability of the resulting clusters. NPs formulated using 0.5 and 1 wt% SPAN® 80 (revealing mean diameters of 223 and 234 nm, respectively) are also found to aggregate over several hours in a 50 mM HCl solution; however, this could be prevented by dialyzing the NPs immediately postproduction to remove surplus surfactant. Dialysis is performed in 10 mM HCl (pH = 2) to ensure colloidal stability. Due to the increased hydrophobicity of the gelatin product consequent to the 4vbc functionalization, the NPs have a maximum cationic ζ-potential of + 23 mV at pH 1, 10 mM salinity (**Figure S5**). Consequently, acidic solutions (10 mM HCl) are employed for particle purification and storage, aiming to suppress any tendency towards aggregation. In the alkaline conditions of pancreatic juices (pH = 7.5–8),[63] the NPs exhibit a ζ-potential of –12 mV (Figure S5), indicating limited long-term colloidal stability. However, since the NPs are delivered directly to the therapeutic site and exhibit rapid degradation in protease-rich environments (**Figure 2**), aggregation or accumulation concerns are mitigated.

NPs fabricated using 10 wt% gel4vbc with a 1:16 aqueous phase to solvent volume ratio and 1 wt% SPAN® 80 are deemed optimal due to reproducibility in production and desirable size for implementation as coating of a nanofiber system. Figure 1B demonstrates the stability of the UV-cured gel4vbc NPs in 10 mM HCl at 20-22°C over four weeks. The NPs show approximately 200 nm in mean diameter with a polydispersity index of 0.2, indicating a narrow particle size distribution independent of the duration of postproduction incubation ($R^2$ values against time equal to 0.05 for NP size and 0.01 for polydispersity). Freeze-drying also does not alter the particle diameter/morphology. Scanning electron micrographs show spherical structures (Figure 1C) with diameters of 227 ± 87 nm (Figure 1D, n = 335), demonstrating comparable sizes to the NPs analyzed in suspension using MA-DLS.

To confirm the removal of diethyl ether, gas chromatography-mass spectroscopic analysis of NPs is performed prior to dialysis, to identify trace amounts of the solvent. No trace of diethyl ether is observed across multiple samples, indicating complete solvent evaporation and successful particle purification (**Figure S6**).

The final particle characterization is performed to determine if the gel4vbc NPs are susceptible to enzymatic degradation in physiological environments (37 °C, pH 7.4), given the amino acidic composition of gelatin. In a control experiment, UV-cured bulk gel4vbc hydrogels (12-14 mg) demonstrate that the UV-induced covalent crosslinking leads to increased resistance to collagenase degradation (**Figure S7**). Complete degradation (100 % mass loss) of an uncured dry sample is measured within 3 h (1 CDU·ml$^{-1}$ collagenase A), while 18 ± 8 % of the UV-cured dry mass was retained after 24 h. Cured gel4vbc NPs (23-25 mg) incubated in a 0.6 CDU·ml$^{-1}$ collagenase solution display over 60 % of mean mass retention after 6 h. On the other hand, insignificant changes in mass are recorded in collagenase-free media, in line with the predominantly enzymatic, rather than hydrolytic, degradability of gelatin (**Figure 2A**).

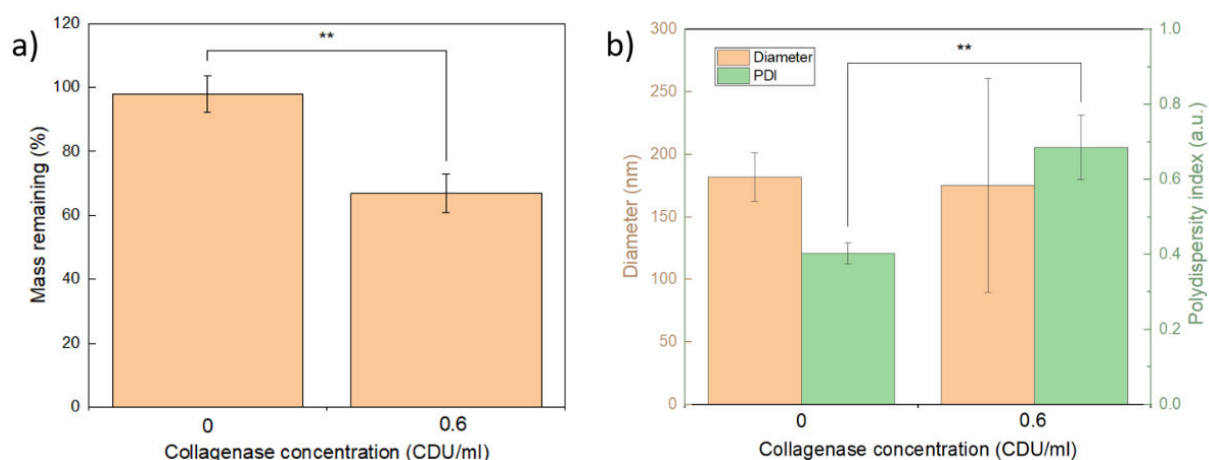

**Figure 2.** Gravimetric (A) and dimensional (B) changes in gel4vbc NPs following incubation (6 h, 37°C) in a Tris buffer containing calcium chloride either without or with collagenase A supplementation (0.6 CDU·ml$^{-1}$). The mass of remaining particles was measured before dispersion into 10 mM HCl to determine size and polydispersity. p-values indicate: ** ≤ 0.01.

Other than gravimetric analysis, Figure 2B depicts the changes in particle size revealed by the aforementioned collagenase degraded NPs (Ø =175 ± 86 nm, PDI = 0.69 ± 0.09). A wider spread of diameters and a significant increase in polydispersity are observed compared to both NPs incubated in enzyme-free media (Ø = 182 ± 19 nm, PDI = 0.40 ± 0.03) and the corresponding freshly synthesized NPs (Ø = 193 ± 24 nm, PDI = 0.17 ± 0.01). The slightly lower NP mean diameters indicate that the NPs undergo surface erosion, in line with the prompt enzymatic cleavage of peptide bonds,[64] with minimal aggregation observed, despite the relatively low value of ζ-potential observed at comparable pH values (Figure S2).

## 2.2. Electrospinning PCL and application of NP-loaded PVA coating.

Electrospun fabrics containing sub-micron PCL fibers were chosen as the basis for the delivery system. This is because the porous structure of the fabric provides a high specific surface area onto which NPs can be coated, providing a homogenous distribution throughout.[65] The innate conformability and tunable chemical properties (biodegradability, hydrophilicity, and biocompatibility) of electrospun fabrics also allow for facile attachment to soft robot surfaces and adhesion in physiological conditions.[66–69]

Electrospinning is performed using a polymer solution of 6 wt% PCL in HFIP, a potential difference of 10 kV, a needle-to-collection plate distance of 10 cm, and an extrusion rate of 1.5 ml h$^{-1}$. Fiber formation is continuous with no bead observed under SEM (**Figure 3A**). The nanofibers have a positively skewed diameter distribution with a modal size of the nanofibers in the range of 300 – 400 nm (22% of the population, **Figure S8A**) and a lognormal mean diameter of 5.81 ± 0.61 (Figure S8B). Positively skewed fiber diameters are obtained when using a low spinning voltage (10 kV),[70] whereby increasing the voltage also yields electrospun fibers with increased thickness creating a Gaussian distribution.

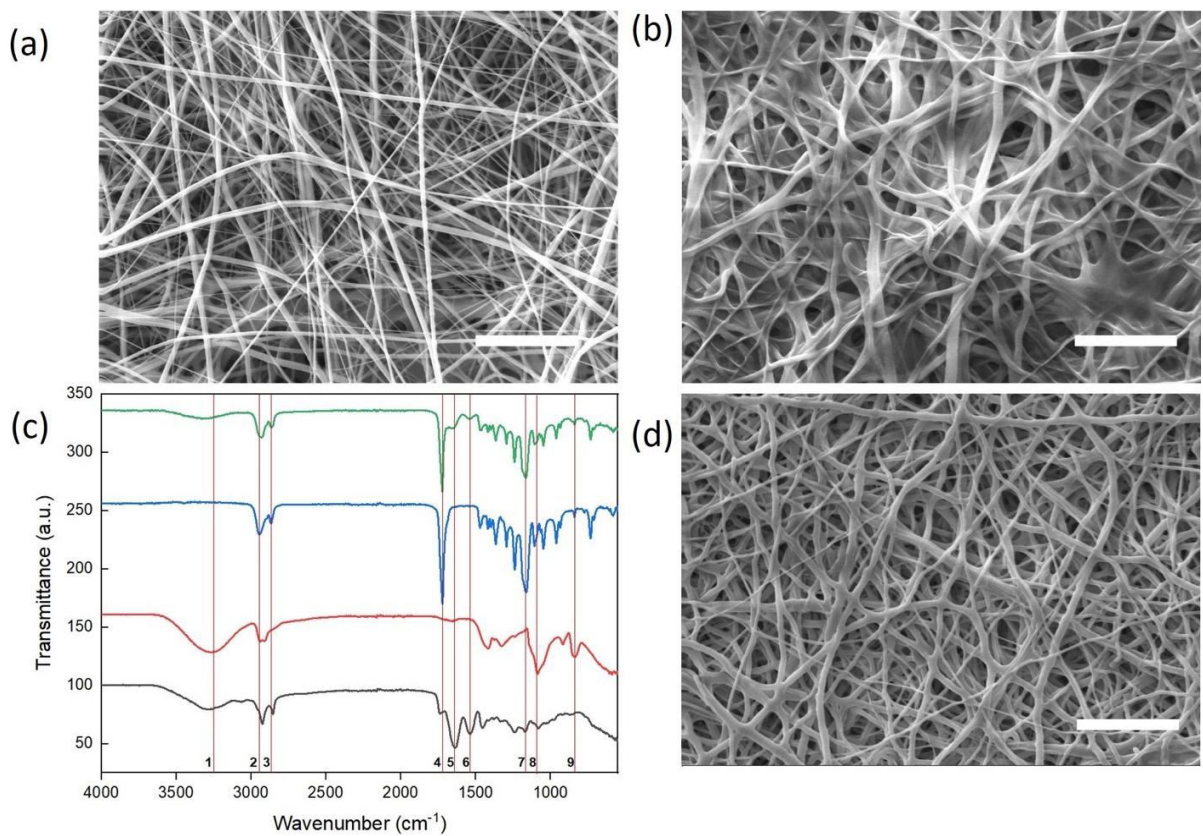

**Figure 3.** Scanning electron micrographs of PCL fabric samples: (A) PCL electrospun fibers and B) NP-coated fabric. C) FTIR spectra of gel4vbc NPs (grey), PVA (red), PCL electrospun fibers (blue), and NP-coated fabric (green). Vertical lines denote FTIR bands associated with chemical bonds characteristic of selected raw materials: 1) 3249 cm$^{-1}$ (O-H stretching), 2) 2945 cm$^{-1}$ (asymmetric CH$_2$), 3) 2865 cm$^{-1}$ (symmetric CH$_2$), 4) 1720 cm$^{-1}$ (C=O), 5) 1640 cm$^{-1}$ (C=O stretching), 6) 1550 cm$^{-1}$ (N-H bending), 7) 1155 cm$^{-1}$ (asymmetric symmetric C-O-C band), 8) 1083 cm$^{-1}$ (C=O stretching and OH bending), 9) 822 cm$^{-1}$ (C-C stretching). D) SEM of the NP-coated fabric after US irradiation (3 W cm$^{-2}$ for 3 min, at 1 MHz 50% duty cycle at zh100 Hz). All scale bars = 10 μm.

Having assessed fiber dimensions, attention focused on subsequent fiber functionalization with NPs. To enable rapid on-demand anticancer delivery, it is hypothesized that an adhesive PVA coating could facilitate effective attachment of anticancer NPs while suppressing their passive diffusion from fiber surfaces. Such a microscale configuration is expected to enable the triggered release of NPs subject to suitable external stimuli due to the positioning of the NPs within the pores of the electrospun fabric. An aqueous solution of PVA (1 wt%) is therefore supplemented with NPs (0.5 wt%), prior to its coating onto the electrospun fabric (0.1 ml·cm$^{-2}$), followed by air drying. PVA is selected as the coating polymer due to its industrial applicability as an adhesive binder and its clinical employability,[71,72] aiming to minimize risks of coating delamination, on the one hand, and ensure cell tolerability, on the other hand.

SEM of the NP-coated fabric indicates that the dried PVA polymer bridges void spaces between the electrospun fibers in the fabric (Figure 3B), in line with the application of the PVA coating on the surface of the fabric. FTIR chemical analysis confirms the presence of both the PVA and the gel4vbc polymers within the sample (Figure 3C). The signals at 3250 cm$^{-1}$ and 1640 cm$^{-1}$, respectively attributed to O-H stretching in PVA, and C=O bond stretching in gelatin, are identified within the spectra of the PCL-based electrospun fabric. FTIR spectra of the front and rear of the NP-coated fabric display signals at identical positions indicating that the PVA and the NPs penetrate the entirety of the fabric (**Figure S9A**), reflecting the relatively low concentration of PVA in the coating mixture. The mat thickness (54 ± 25 µm) is greater than the focal length of the FTIR beam (2 µm), indicating that only the surface level of the front and rear of the fabric are analyzed.

Application of US to the fabric (3 W cm$^{-2}$ for 3 min, at 1 MHz 50% duty cycle at 100 Hz) shows the removal of the NP-loaded PVA coating bridging between fibers (Figure 3D). The removal of the NP-loaded PVA coating from respective fabric is further supported by FTIR analysis (Figure S9B). The signals at 3250 and 1640 cm$^{-1}$ are reduced initially when the polymer-coated fabric is incubated in an aqueous environment for 1 h, with further reductions upon US application. This indicates that the PVA and NPs are solubilized in the aqueous medium allowing fiber dislocation, an effect that can be enhanced through US irradiation. On the other hand, US exposure is found to increase the average fiber diameter. The US-induced pressure perturbations cause a relaxation of the polymer chains enabling access of water molecules.[73,74] The water interacts with the polar regions within PCL, breaking or inhibiting the secondary interactions between, and structural rigidity of, polymer chains.[75,76] The nanoparticle distribution remains positively skewed with an increased count in the modal diameter range at 300 – 400 nm (25% of the fiber population, Figure S8C). However, the lognormal mean of US-treated fabric is found to be statistically increased (p-value: 0.04) to 6.02 ± 0.52 (Figure S8D). Fiber fusing can also be observed in the US-treated sample. PCL nanofibers can undergo morphological changes (curvilinear deformation, fusing, and beading) at temperatures as low as 36°C,[77] despite bulk PCL having a melting temperature of ~60°C. The US irradiation at 3 W·cm$^{-2}$ for 3 min causes the bulk temperature within the aqueous media (1 ml) to increase to approximately 35°C (**Figure S10**). Furthermore, with the PVA layer filling the void spaces within the fabric and bridging adjacent fibers, the likelihood of thermally induced fusion is increased.

The mechanical properties of the electrospun fibers are also investigated. Samples of 4 cm x 1 cm are subjected to a 5 N load to obtain force-displacement data from which stress-strain curves were generated (**Figure S11**). The addition of the NP-loaded PVA coating to the PCL fabrics significantly increases their ultimate tensile strength, with a maximum load of 1289 ± 133 kPa in the pristine electrospun sample and 2409 ± 831 kPa in the corresponding fabric. The improved mechanical properties reflect the adhesive bonding effect of the PVA coating on the fibers, yielding a greater mechanical resistance to failure.[78] PBS hydration and US irradiation do not affect the tensile strength with maximum stresses of 2494 ± 720 kPa and 2669 ± 573 kPa, respectively. As demonstrated in Figure S9B, a substantial amount of the PVA remains within the fabric structure after 1 h of incubation, meaning the PVA continued to provide reinforcement. However, the US has been shown to induce removal of the PVA from the fabric, suggesting a decrease in the ultimate tensile strength. The minimal variation in mechanical strength is therefore attributed to the US-induced fiber fusion, as previously observed in Figure 3D. Consequently, it is possible to accomplish US-triggered release of the NP-loaded PVA coating without compromising the overall mechanical behavior of the underlying fabric.

Despite the increased mechanical strength revealed by the dry fabric, no variation in fracture strain is measured, indicating minimal effect of the NP-loaded PVA coating (**Figure 4**).

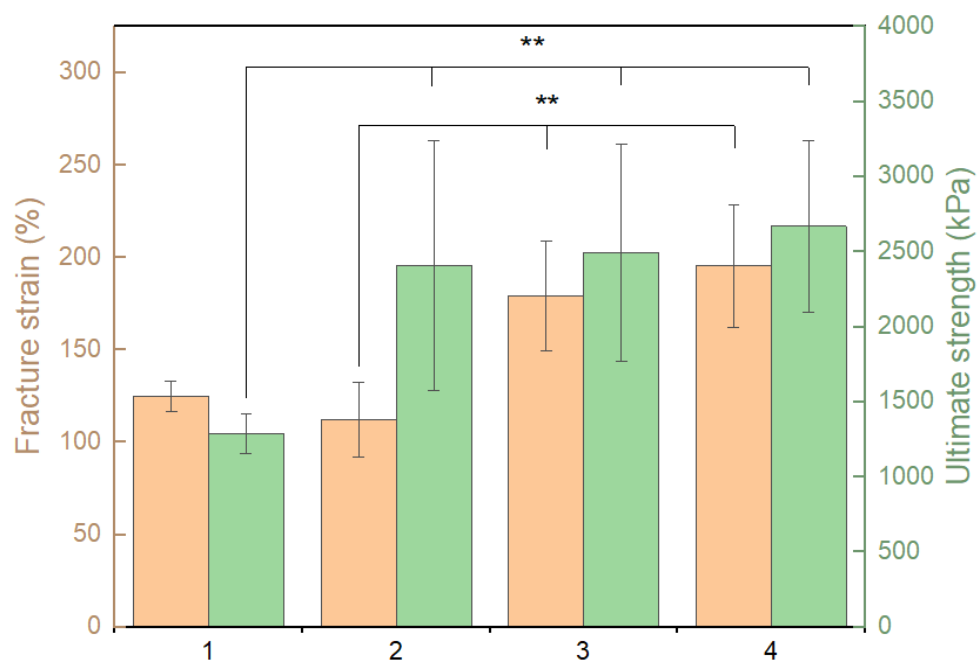

**Figure 4.** The fracture strain (orange) and ultimate strength (green) of PCL fabric samples: 1 – PCL electrospun fabric, 2 – NP-coated fabric, 3 –NP-coated fabric following 1 h in PBS, 4 – NP-coated fabric following 1 h in PBS and US application (3 min at 3 W cm$^{-2}$ at 1 MHz, in pulsed mode with a duty cycle of 50% and at 100 Hz). In all figures, error bars indicate the standard deviation (n = 3), and p-values indicate: ** ≤ 0.01.

However, upon incubation in PBS, the elongation ($\varepsilon$) is significantly increased from 112 ± 20 % to 179 ± 30 %, while no further change is observed upon US application ($\varepsilon$ = 196 ± 33 %). The increased extensibility is attributed to the bound water acting as a plasticizer, softening the PCL and PVA chains and allowing them to stretch.[75,76,79]

### 2.3. Ultrasound-triggered delivery capability of gel4vbc NPs.

Passive NP release from the NP-coated fabric is assessed by measuring the solution turbidity at 390 nm when incubated in 10 mM HCl. The extent of NP diffusion from the fabric is affected by the PVA concentration in the coating-forming solution (**Figure 5A**). When PVA-free NP-loaded aqueous suspension is applied to the fabric, faster NP release is observed *in vitro*, corresponding to 54 ± 11 % NP release within 15 min, increasing to 86 ± 1 % after 24 h. When low concentrations of PVA (0.1 wt%) are introduced in the coating-forming solution, the NP release after 15 min fabric incubation in PBS is decreased to 43 ± 6 %. The increased NP retention following fabric coating with the PVA-supplemented NP-loaded solution is likely due to the PVA-induced binding effect, limiting the particle diffusion and dislocation from the fabric structure. This effect is enhanced when coating solutions with increased PVA concentration (up to 0.5 wt%) are applied to the fabric, corresponding to a 17 ± 7% NP release after 15 min and 35 ± 3 % after 24 h. When coating solutions with a PVA concentration above 0.5 wt%, i.e. 1 wt% PVA, are employed, there is no significant increase in NP passive retention within the corresponding fabric, so that 7 and 33 % of NPs are released after 15 min and 24 h, respectively.

In addition to passive diffusion, the PVA concentration in the fabric-coating solution also impacts the ultrasound-triggered release of NPs from the corresponding fabric. Figure 5B shows similar trends in US-triggered NP release following either 15- or 60-min sample incubation in aqueous media. The NP release from the fabric is found to increase when coating solutions with increasing PVA concentrations (i.e. from 0 to 0.5 wt%) are applied. As such, the maximum US-triggered NP release is observed when 0.5 wt% PVA-supplemented solution is coated onto the fabric, corresponding to 74 ± 17 % and 76 ± 4 % NP release after US irradiation for 15 and 60 min, respectively. The aforementioned increase in NP release from fabrics prepared with coating solutions supplemented with increasing PVA concentration may appear counterintuitive, due to the aforementioned PVA-induced binding capability, as observed when incubation of the fabric is performed in PBS in the absence of US. On the other hand, the fabric coatings prepared with higher PVA concentration induce increased NP retention within the fibrous structure up to the point of the US application. The summation of the US-free passive and US-triggered NP release profiles (Figure 5C) provides supporting evidence that the US agitation enables the release of all remaining NPs.

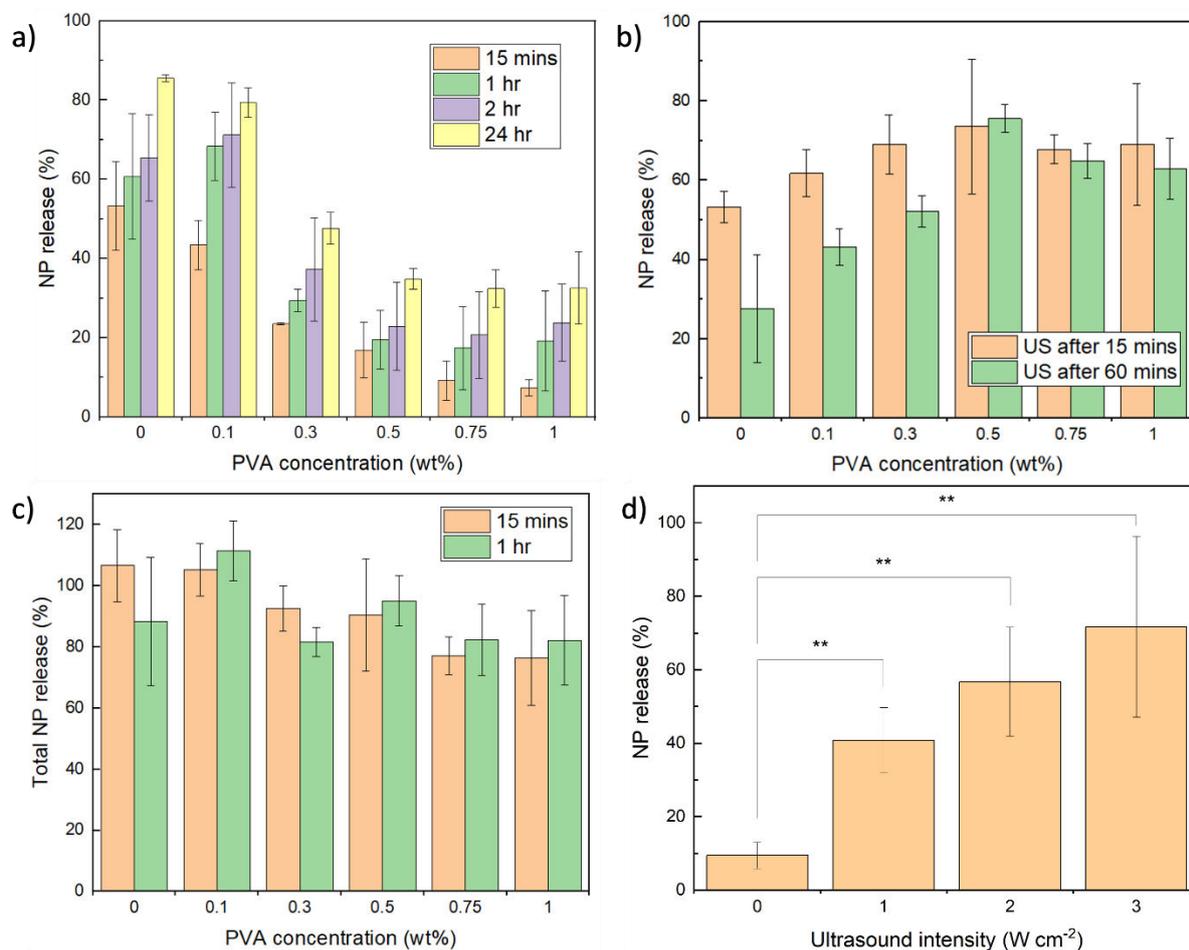

**Figure 5.** A) Passive release of NPs from PCL electrospun fabrics coated with NP-loaded solutions supplemented with varied PVA concentration. The NP release from the fabric was quantified after incubation for 15 min (orange), 1 h (green), 2 h (purple), and 24 h (yellow). B) NP release from the aforementioned fabrics when US was activated after incubation for either 15 or 60 min (orange and green, respectively). C) Summation of passive and US-triggered NP release from fabrics coated with aqueous solutions loaded with NPs (1 wt%) and varying PVA concentrations. US agitation occurred after 15 (orange) or 60 (green) min incubation in PBS. D) NP release from NP-coated fabrics hydrated for 15 min before the US application at varying intensities. All samples were incubated in 10 mM HCl at 37°C, and US was applied for 3 min at 3 W cm$^{-2}$ at 1 MHz, in pulsed mode with a duty cycle of 50% and at 100 Hz, unless otherwise stated. In all figures, error bars indicate the standard deviation (n = 3), p-values indicate: ** $\leq$ 0.01.

All samples coated with the PVA-supplemented (0–0.5 wt% PVA) NP-loaded solutions reveal above 90 % of total release. The 100% NP capacity is obtained by taking the average turbidity of the NP and PVA solutions prior to the addition to PCL mats. As such, the total NP release values above 100% are attributed to the variation in the maximum loading value. PVA concentrations above 0.5 wt% in the NP-loaded coating solution prove to inhibit NP release from the corresponding fabric, whereby a total NP release (US and 15 min hydration) of 77 and 76 % are recorded with samples supplemented with 0.75 wt% and 1 wt% PVA, respectively.

The effect of the US intensity on NP release is investigated by applying US at varying intensities (0–3 W cm$^{-2}$) for 3 min after 15 min incubation in 10 mM HCl. All samples subjected to US demonstrate a significant increase in NP release compared to the passive release (Figure 5D). The average NP release is found to increase with US intensity, from 41 ± 9 % at 1 W cm$^{-2}$ to 72 ± 25 % at 3 W cm$^{-2}$, supporting the effect of local ultrasonic cavitation-based mechanical force on the release capability of the fabric.[80]

**2.4. US-triggered release of anticancer surrogate incorporated with soft robotic tentacles.**

Passive and US-triggered release of albumin is presented in **Figure 6A-B**. Albumin loading of the UV-cured NPs is achieved by particle swelling in a 1 wt% albumin solution. The suspension is freeze-dried and the retrieved NPs dispersed in DI H$_2$O prior to centrifugation and dispersion in an aqueous solution supplemented with 0.5 wt% PVA. This protocol enables the loading of 1.29 ± 0.18 mg ml$^{-1}$ of albumin into a 1 wt% solution of NPs, corresponding to a loading efficiency of 3.87 %, as ascertained using a calibration plot (**Figure S12**).

Loading of albumin into NPs (1 wt%) and their coating within the fabric delay the passive release from the fabric compared to the case where NP-free albumin is coated onto the fabric. Within 15 min of the fabric submersion in PBS, 84 ± 10 wt% of the total NP-free albumin has been released into the supernatant, rising to 98 ± 3 % in 30 min. This contrasts with the 65 ± 1 % release that is recorded when albumin-loaded NPs are coated on the fabric and the resulting fabric incubated for 15 min. The controlled release capability of the albumin-loaded NPs is observed over a 180 min duration of incubation, at which point 82 ± 2 wt% albumin is released. US agitation of the fabric coated with albumin-loaded NPs after 15 min hydration shows an additional mean burst release of 34 wt% of the albumin payload, equating to a total albumin release of 98 ± 5 wt%. The US-triggered albumin delivery is also effective after prolonged hydration, whereby US activation after 60 min of incubation shows a statistically significant 20 % increase in albumin release. The release of albumin from the NPs is driven by sonomechanical agitation. Figure 5 demonstrates that the nanoparticles remain structurally intact following release via ultrasound (US) irradiation, as evidenced by increased solution turbidity after ultrasound exposure. The particle-forming gelatin molecules are crosslinked via newly-formed, UV-induced carbon-carbon (C–C) bonds, which are characterized by high dissociation energies and minimal biodegradability risks,[81] in line with the structural stability of the UV-cured gel4vbc material even after heating at 100°C for 15 minutes (Figure S3).

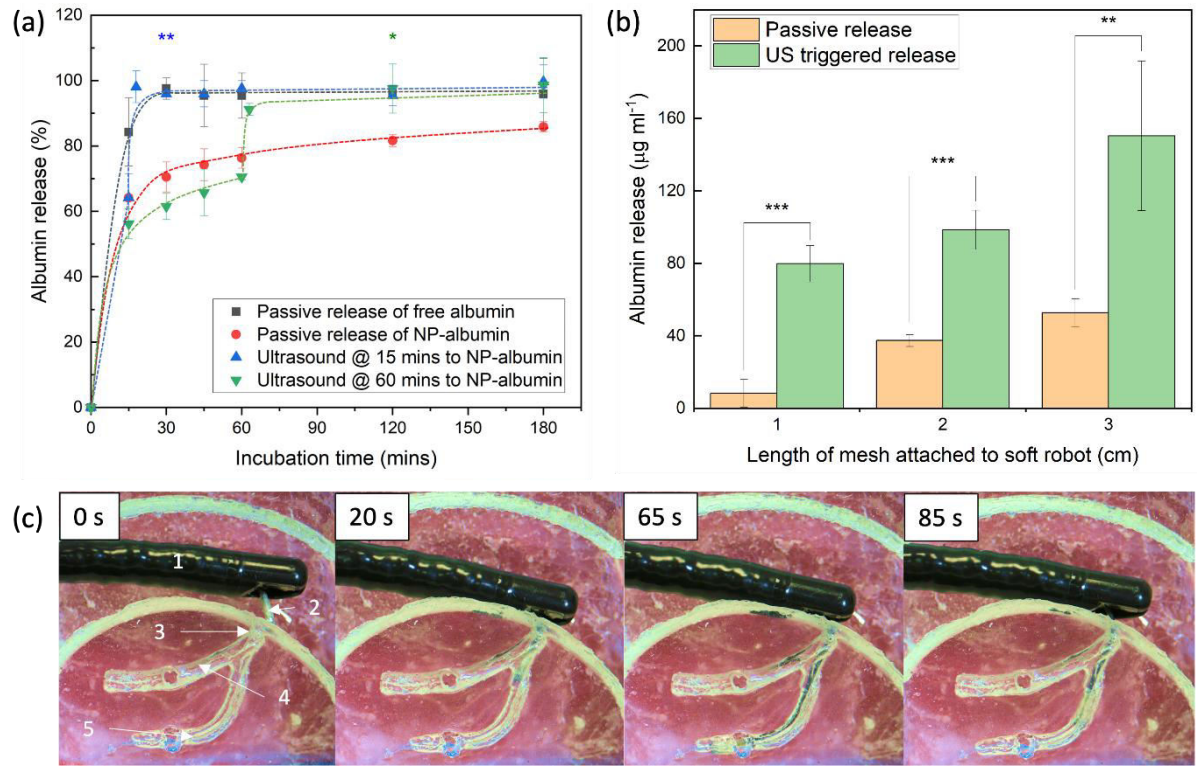

**Figure 6.** (A) Passive and US-triggered albumin release measured following incubation (PBS, 37°C) of the PCL fabric coated with the PVA-supplemented aqueous solution (0.5 wt% PVA) containing either 0.1 wt% free albumin (grey) or 0.1 wt% NP-loaded albumin (red). US was applied after either 15 (blue) or 60 min (green) of incubation in PBS. Statistical significance is determined between the US-triggered release and the passive release measured from the NP-loaded fabric. (B) Passive (green) and US-triggered (orange) albumin release from NP-loaded fabric attached to a magnetic tentacle robot (MTR). Fabric strips (1 – 3 cm x 1 cm) were coated (0.1 ml cm$^{-2}$) with PVA-supplemented solutions (0.5 wt% PVA) containing 0.1 wt% albumin-loaded NPs. US was always applied for 3 min at 3 W cm$^{-2}$ at 1 MHz, pulsed mode with a duty cycle of 50% at 100 Hz. In all figures, error bars indicate the standard deviation (n = 3), p-values indicate: * ≤ 0.05, ** ≤ 0.01, *** ≤ 0.001. (C) Images of PCL fabric coated with albumin-loaded NPs attached to a soft robotic tentacle during magnetic manipulation out of a duodenoscope into a phantom pancreatic duct and the retraction process. Labels indicate: 1 – duodenoscope, 2 – mat coated MTR, 3 – major duodenal papilla, 4 – bile duct, 5 – pancreatic duct. Scale bar = 2 cm applies to all images.

Collectively, these observations indicate that the NPs withstand both bulk heating and the mechanical forces generated during ultrasound exposure. Consequently, it is hypothesized that a synergistic combination of acoustic radiation forces, which transiently deform the hydrogel network, and acoustic streaming, which promotes internal fluid motion,[82] drives the movement and subsequent release of encapsulated molecules from the nanoparticles.

Moreover, brief heating of the sample to 35°C for 3 minutes is not expected to adversely affect drug bioactivity, given mild hyperthermia (40–43°C for approximately 1 hour) is commonly employed to enhance the efficacy of chemotherapeutics against cancer.[83]

To demonstrate integration with soft robotics, the NP-coated fabric is loaded onto the tip of a magnetic tentacle robot (MTR) that is drawn out of a protective duodenoscope (**Figure 6C**). Fabric attachment is achieved by fixing one end of the strip on top of the tentacle (using an aqueous solution of PVA, 30 wt%), wrapping the strip around the tentacle, and fixing the other end of the strip to the underlying fabric layer. Figure 6C describes the introduction of the duodenoscope and subsequent magnetic manipulation of the fabric-coated MTR into a phantom pancreatic duct. The additional fabric layer does not interfere with the soft tentacle motion out of the duodenoscope or the magnetic navigation into the pancreatic duct of an Ecoflex™ phantom. The time stamps reported in Figure 6c show that the MTR can be maneuvered quickly within soft and complex networks. After positioning the duodenoscope adjacent to the major duodenal papilla, the tentacle is guided past the bile duct into the pancreatic duct and retracted back into the duodenoscope in under 90 seconds. The duodenoscope sheath protection from the wet physiological environment and the rapid magnetic navigation of the tentacle to the target site location will limit premature release of chemotherapeutics from its fabric-coated tip, maximizing release upon US irradiation.

All samples exhibit a statistically significant increase in albumin release within the phantom pancreatic duct, compared to those incubated in the duct for five minutes without US agitation (Figure 6B). All fabrics remained securely attached to the MTR throughout the entire procedure, indicating that the MTR response to US exposure was not associated with significant increase in solution temperature, given that PCL has a melting temperature of approximately 60°C. Furthermore, the design approach used to build the NP-coated fabric enables prompt variation in albumin delivery by changes in the area of the strip applied to the MTR. For instance, while a 1-cm$^2$ MTR-tipped fabric is found to release 80 µg·ml$^{-1}$ of albumin under US agitation, nearly double this release is accomplished in the same conditions when a larger strip (3 cm x 1 cm) is applied to the MTR. The non-linear scaling of delivery with surface area is due to the mat coiling over itself, inhibiting drug release from the lower layers. The increased albumin delivery with the larger fabric indicates the potential for facile control over the drug dose by varying the surface area attached to the robot.

## 3. Discussion

This study presents a novel integrated design to achieve localized, on-demand drug release for pancreatic cancer applications, by leveraging the biodegradability of gelatin-based NPs,[84] the conformability of electrospun fabric[85] and the magnetic responsivity of a soft robotic endoscopic platform.[86] The ultrasound (US)-responsive drug carrier system comprises an electrospun poly(ε-caprolactone) nanofibrous fabric that is functionalized with albumin-loaded gel4vbc NPs via PVA coating binder. This fiber-based construct is mounted on a Magnetic Tentacle Robot (MTR) that can be navigated through an endoscope into the pancreatic duct.[19] Once positioned at the target site, the nanoparticle release is activated by unfocused US, allowing fine spatiotemporal control over therapeutic release with minimal safety risks, given the biocompatibility and biodegradability of gel4vbc.[47] In the absence of US, the fabric only slowly releases its payload, indicating that the presence of the NPs and PVA binder enables effective drug retention within the construct until US irradiation. In addition to the on-demand release capability, this design can be tailored to accomplish tuneable release profiles. Increasing the concentration of the PVA binder in the coating leads to higher nanoparticle retention and a greater fraction of US-triggered release. Similarly, the system allows dosage customization by altering the surface area of the MTR-coating fabric, whereby a threefold increase in fabric size contributes nearly doubled the amount of drug release under the same ultrasound exposure. Such a level of release control in space (confined to the fabric location), time (on-demand release via ultrasound), and dosage, surpasses that of earlier local delivery systems, which generally rely on passive or in situ triggers and are not easily adjustable once deployed.[87]

Regarding delivery precision and controllability, the ultrasound-responsive MTR-coated fabric introduces clear advantages. Unlike controlled-release implants or injectable gels that continuously diffuse drugs, the described system retains its therapeutic payload until the clinician activates it at the tumor site. This feature is key aiming to minimize risks of premature drug leakage and side toxicity during device navigation,[88] and concentrate the dose release to the tumor region upon external activation. In this context, the application of US is particularly appealing, given its established clinical use and relative safety and the lack of interference with the magnetic field used to enable device navigation to the tumor site.[89] The fabric integration with a soft MTR therefore further enhances spatial precision: it enables gentle, minimally invasive access deep into the pancreatic ductal network, potentially reaching tumors that conventional rigid tools or systemic therapy might not effectively treat.[19] The proposed system is the first to apply such principles in a pancreatic cancer context, combining a US-responsive fabric with an endoscopic robot. This unique combination positions it as a highly novel approach relative to prior studies,[90] which have typically implemented either stimulus-responsive materials or advanced endoscopic devices, but not their integration into a single multifunctional device.

Despite its promise, this study has limitations. Currently, validations are confined to *in vitro* and phantom models, with no *in vivo* or preclinical tumor studies reported. Consequently, the efficiency and safety of US-triggered release in real pancreatic tissues, which may cause attenuation or scattering of ultrasound, remain to be demonstrated. Additionally, albumin served as a surrogate payload, and its solubility and release characteristics may differ from those of smaller, clinically relevant chemotherapeutics.[91] Lastly, scaling the fabric's surface area to increase drug dosage showed a practical upper limit due to non-linear drug release caused by fabric overlap. Strategies like deploying multiple smaller fabric patches or optimizing fabric geometry may prove key to overcome this challenge. Future studies should therefore prioritize loading clinically relevant anticancer agents into the nanoparticles and conducting comprehensive biocompatibility testing in appropriate environments *in vitro* and animal models, together with investigations on sterilisation and chemical analysis of any extractable and leachable. Optimizing ultrasound parameters and seamless integration with medical robotic systems will be crucial to transitioning this promising drug delivery approach into preclinical and clinical applications.

## 4. Conclusion

A novel drug delivery system is presented for the facile integration of soft robotics systems with on-demand and localized therapeutic capabilities. By adapting a nanoemulsion-based NP fabrication strategy, we have created size-tunable UV-cured NPs, with traceless solvent phase post-processing, prolonged shelf-life colloidal stability and delayed enzymatic degradability. Using NP-loaded PVA-supplemented aqueous solution, we demonstrate the first example of ultrasound-triggered delivery of anticancer NPs from PCL electrospun nanofibrous fabrics coupled to an MTR. Pulsed mode US (50 % duty cycle, 100 Hz repetition frequency) applied for 3 min at 1 MHz with intensities between 1 – 3 W cm$^{-2}$ successfully triggers an additional 32 – 63 % NP release from the fabric. Varying the PVA concentration in the fabric coating solution provides further control of the release profile, with higher PVA concentrations (0.5 - 1 wt%) enabling enhanced NP retention and increased triggered release. Loading albumin into NPs offers a simple means to equip the fabric with controlled drug delivery, which is key to ensuring anticancer treatment efficacy. 64 wt% albumin release is measured after 15-min incubation *in vitro* of the NP-coated fabric, as opposed to 84 wt% release from NP-free albumin-loaded samples. The demonstrated capability to integrate a NP-coated fabric with a magnetically controlled tentacle robot enables rapid navigation into the pancreatic duct of a duodenum phantom, to enable highly localized, on-demand (US-triggered) anticancer release. This design strategy also offers as easy route to regulate the drug dosage by controlling the surface area of the NP-loaded fabric adhered to the robot, so that is possible to achieve a nearly two-fold increase in albumin release (80 à 150 µg·cm$^{-2}$) through a three-fold increase in fabric size (1 à 3 cm$^2$). As such, our design offers control over drug dosage by cutting the fabric into different shapes/sizes, while the flexibility and small thickness (~54 µm) allow the fabric to conform to complex device topographies, widening the scope of potential robotic implementation. This design is a proof-of-concept for US-responsive fabrics for endoluminal technologies, opening a novel avenue to pancreatic tumor treatment.

## 5. Experimental Section/Methods
### 5.1. Materials
PVA ($M_w$ 146-186 kDa, 99 % hydrolyzed), PCL ($M_n$ 80 kDa), gelatin from porcine skin (high bloom ~300), 4VBC (90%), 2,4,6-trinitrobenzene sulfonic acid solution (TNBS), diethyl ether, collagenase A (from *Clostridium histolyticum*), phosphate buffered saline (PBS) tablets, tris(hydroxymethyl)aminomethane (Tris), absolute ethanol, silicone oil AR 200, albumin from chick egg white (Grade 2), polysorbate 20 (TWEEN® 20) and sorbitan monooleate (SPAN® 80) were purchased from Merck (Darmstadt, Germany). Calcium chloride (dried, 97%) was purchased from Alfa Aesar (Massachusetts). Sodium bicarbonate (NaHCO$_3$) and 2-hydroxy-1-(4-(2-hydroxyethoxy)phenyl)-2-methylpropan-1-one (Irgacure I2959) were purchased from Fluorochem Ltd (United Kingdom). 1,1,1,3,3,3-hexafluoroisopropanol (HFIP) and triethylamine (TEA) were purchased from Apollo Scientific (United Kingdom). Hydrochloric acid (6 N) was purchased from Fisher (Massachusetts).

### 5.2 Ultrasound treatment
Ultrasound (US) treatment was performed using a 5 cm² collimating transducer connected to an SP300 sonoporation system (SONIDEL, Dublin), operated in unfocused mode. Aquaflex® ultrasound gel pads were employed as spacers between the well plate and transducer to maintain a consistent distance between the sample and US source. Unless otherwise specified, all samples were irradiated under the following conditions: 3 minutes duration, intensity of 3 W cm$^{-2}$, frequency of 1 MHz, and pulsed mode operation at a repetition rate of 100 Hz with a 50% duty cycle. The selected frequency of 1 MHz provides an optimal penetration depth for deep tissue ultrasound therapy,[83] ensuring effective access to the pancreatic capillaries. All other parameters were optimized specifically for interaction with the electrospun material to achieve efficient nanoparticle release without exceeding physiological temperatures.

## 5.3. Gelatin functionalization with 4VBC and chemical characterization

Native gelatin was dissolved (10 wt%) into PBS (pH = 7.4) by heating to 45°C and mixing vigorously for 1 h. 1 wt% of TWEEN 20 was added dropwise, followed by rapid addition of 4VBC and TEA, both at 25 times molar excess with respect to the molar content of gelatin lysine groups. The solution was left to mix for 5 h before precipitation into 10-times volume of absolute ethanol. The precipitate was removed the following day and left to dry in a fume hood for 72 h. To quantify the degree of functionalization, samples were analyzed via the TNBS assay.[57] $NaHCO_3$ (4 %) and TNBS (0.5 %) solutions (both 1 ml) were added to either gel4vbc or native gelatin samples (11 mg) and heated to 40°C for 4 h. Samples were then dissolved by adding 3 ml of 6 N HCl and heated to 60°C for 1 h. After cooling to room temperature, the solutions (5 ml) were diluted in equal volumes of distilled water (DI $H_2O$), and unreacted TNBS was removed through diethyl ether extraction. Samples were heated to 60°C to evaporate trace diethyl ether. 5 ml of the remaining solution was diluted with 15 ml of DI $H_2O$. The absorbance was measured at 364 nm and placed into **Equation 1**:

(Equation 1)    $\text{moles(Lys)/g(protein)} = (2 \cdot A_{346} \cdot 0.02)/((b \cdot x \cdot (1.46 \cdot 10^4 \text{ L cm/mole})$

where 2 is the dilution factor, 0.02 is the sample volume (liters), $1.46 \times 10^4$ is the molar absorption coefficient for 2,4,6-trinitrophenyl lysine, b is the cell path length (1 cm), and x is the sample weight in grams. The degree of functionalization was calculated according to **Equation 2**:

(Equation 2)    $F = (1 - Lys(Gel4vbc)/Lys(Gel)) \cdot 100$

where F is the degree of functionalization (%), *Lys(Gel4vbc)* and *Lys(Gel)* refer to the lysine molar content in 4VBC-reacted and native gelatin, respectively.

Confirmation of UV-induced crosslinking was achieved using a Differential Scanning Calorimeter 4000 (Perkin Elmer, Massachusetts) to observe changes in the onset of the gel-sol transition of bulk gels. Native gelatin and gel4vbc gels (cured and non-cured) were freeze-dried before 3 – 5 mg samples were evaluated between 5 - 100°C, with a heating rate of 5°C min$^{-1}$.

## 5.4. Nanoparticle manufacture

Gel4vbc nanoparticles (NPs) were fabricated using a nanoemulsion-based method. Briefly, gel4vbc (10 wt%) was dissolved into 10 mM HCl, at 45°C with mixing. Diethyl ether (20 ml) was added to a flask and mixed vigorously while 1 wt% of SPAN® 80 was added dropwise. The gel4vbc solution (1.3 ml) was then added and the solution immediately homogenized for 2 min (8,000 rpm). The solution was left to mix until the bulk diethyl ether had evaporated away, leaving a thick NP paste. The NPs were dispersed into a 10 mM HCl solution containing 1 wt% I2959 and UV-cured for 30 min (Chromato-Vue C-71, Analytik Jena, Upland, CA, USA), with continued vigorous mixing. The creamy solution was then added to a 100 kDa molecular cutoff cellulose dialysis membrane and placed into a 2 L bath of 10 mM HCl and stirred for 48 h. The dialysis medium was replaced with fresh 10 mM HCl after 16 and 24 h. Albumin loading of the NPs was performed by dispersing freeze-dried NPs into a 10 mM HCl solution containing albumin (1 wt%) and mixing for 24 h. The solution was freeze-dried before dispersion in DI H2O and centrifugation at 36,000 x g for 30 min. The supernatant was removed, and the NP pellet was redispersed in a PVA (0.5 wt%) solution.

## 5.5. Nanoparticle characterization

*Sizing and colloidal stability* : Size and polydispersity measurements were taken using multi-angle dynamic light scattering (MADLS) at 20°C, with a Zetasizer Advance Ultra (Malvern, United Kingdom). Gel4vbc NPs diluted 100 times in 10 mM HCl were sized using a 4 mW, 633 nm laser with measurement angles 45°, 90°, and 173°. Between measurements, samples were stored at 20°C in 10 mM HCl.

Scanning electron micrographs were used to size freeze-dried NPs. The dry NPs were spread onto a carbon stub before gold coating, 5 nm. Samples were imaged using an EVO MA15 SEM (Zeiss, Oberkochen) at 15,000-, 10,000-, and 5,000-times magnification. Image processing was performed using Fiji image processing software.

*Enzymatic stability*: Collagenase A was dissolved (0.6 CDU·ml$^{-1}$) in 50 mM tris buffer (pH = 7.4) containing 0.36 mM calcium chloride. The collagenase solution (1 ml) was added to 20 mg of freeze-dried gel4vcb NPs and mixed vigorously. The solution was incubated under gentle mixing at 37°C for 6 h. Samples were freeze-dried; the remaining mass was weighed before dispersion into 10 mM HCl and sizing using MADLS.

*Zeta-potential* : Upon completion of the gel4vbc NP dialysis, the bulk liquid was replaced with DI H$_2$O and dialyzed for a further 1 h. The NPs were then diluted 1:9 in the following buffers (pH adjusted using 0.1 M HCl or 0.1 M NaOH) at 11.1 mM: hydrochloric acid–potassium chloride (pH 1), citrate-phosphate (pH 3, 5), Tris (pH 7, 9), phosphate-sodium hydroxide (pH 11). Z-potential measurements were then performed using a Zetasizer Advance Ultra, at 20°C. Scanning electron micrographs were used to size freeze-dried NPs. The dry NPs were spread onto a carbon stub before gold coating (5 nm). Samples were imaged using an EVO MA15 SEM (Zeiss, Oberkochen) at 15,000-, 10,000-, and 5,000-times magnification. Image processing was performed using Fiji image processing software.

*Gas chromatography mass spectroscopy (GC-MS)*: Measurements were performed using a GC-2010 with GCMS – QP2010 mass spectrometer (Shimadzu, Kyoto) equipped with a Supelco SLB5 MS capillary column, with a length of 30 m, an internal diameter of 0.25 mm, and a film thickness of 0.25 μm. Immediately post-curing, gel4vbc NPs were placed into a closed vial and transferred into the GC-MS. The sample was extracted using a 250 μl injection head space, and injected into the column at 250°C, with an argon carrier gas flow rate of 1.22 ml min$^{-1}$ and a linear velocity of 39.9 cm sec$^{-1}$.

## 5.6. Electrospinning and NP coating of PCL fabrics

PCL (6 wt%) was dissolved into HFIP under mixing for 2 h and left for a minimum of 24 h. Electrospinning was performed using an L1.0 Aerospinner (Areka, Istanbul). The polymer solution was driven with a pump speed of 1.5 ml h$^{-1}$, using a needle-to-collector plate distance of 10 cm, and a collector-needle potential difference of 10 kV. Spinning was performed for 1 h over a 30 cm length and 16 mm s$^{-1}$ homogenization rate with electrospun fibers being deposited on metal foil to form a thin, self-supporting fabric structure.

Prior to coating, the electrospun fabric was submerged into DI H$_2$O to enable removal from the collection foil, prior to air-drying in a fume hood for 24 h and cutting to the desired shape. Gel4vbc NPs (1 wt%) and PVA (0–1 wt%) were mixed in a 10 mM HCl solution and the resulting mixture added to the PCL electrospun fabric at a ratio of 0.1 ml cm$^{-2}$ to generate the NP-coated fabric.

## 5.7. Characterization of NP-coated electrospun fabric

*Fiber dimensions*: Fiber diameter measurements were made using scanning electron microscopy (SEM). Preparation and use of the SEM were performed as previously described. Fiji image processing software was utilized to analyze fibers individually.

*Fourier transform infrared spectroscopy (FTIR)*: Measurements were performed using a Spectrum One FT-IR Spectrometer (Perkin Elmer, Massachusetts) with a Golden Gate™ attenuated total reflectance accessory (Specac Ltd, United Kingdom). Samples were analyzed between 4000 – 550 cm$^{-1}$ with a scan speed of 0.5 cm s$^{-1}$ and a 4 cm$^{-1}$ resolution.

*Mechanical property testing* : Tensile testing was performed using an Instron 3365 Universal Testing Machine (Instron, Massachusetts). Strips of PCL electrospun fabric (1 cm x 2 cm) were subjected to a 5 N load and 5 mm·min$^{-1}$ extension rate to obtain force-displacement graphs. Dry NP-coated fabrics were also tested alongside NP-coated fabrics recovered following 1-hour incubation in PBS (pH 7.4, 37°C) and US exposure. Stress-strain plots were generated from the force-displacement data, so that the ultimate tensile strength and fracture strain values were obtained.

## 5.8. NP and albumin release from NP-coated fabric

NP-coated fabrics (1 x 1 cm) were prepared as previously described. To determine the turbidity of 100% NP release, 0.1 ml of the NP-loaded PVA solution was dried and dispersed into 1 ml of 10 mM HCl, before measuring the absorbance at 300 nm. The NP-coated fabrics were then incubated in PBS (1 ml, pH 7.4) at 37°C and gently mixed for up to 24 h. To observe passive release, the supernatant was extracted and the turbidity measured throughout the incubation period.

To quantify the US-triggered release after either 15 or 60 min of PBS incubation, the fabric was placed into fresh medium and irradiated for 3 min. The turbidity of the supernatant (0.3 ml) was analyzed by measuring the absorbance at 390 nm using a Varioscan LUX multimode microplate reader (Thermo Fisher, Massachusetts). Similarly, albumin release was determined using the same protocol; however, the albumin content was quantified by measuring the absorbance at 280 nm, using a NanoDrop 2000 spectrophotometer (Thermo Fisher, Massachusetts).

## 5.9. Integration of NP-coated electrospun fabric with the magnetic tentacle robot and demonstration of US–triggered delivery in a phantom

To validate the proposed approach, an MTR was fabricated by blending Dragon Skin 30 (Smooth-On Inc, USA) with 5 μm neodymium-iron-boron (NdFeB) microparticles in a 1:1 mass ratio. The MTR had a cross-sectional diameter of Ø = 1.82 ± 0.04 mm, and each sample had a length of 60 ± 1 mm. Subsequently, the MTR underwent exposure to a saturating uniform magnetic field of 4.644 T (ASC IM-10-30, ASC Scientific, USA) along its axis.[58] The drug release fabric was prepared by coating the PCL electrospun sample with a PVA solution (1 wt%) containing albumin-loaded NPs (0.5 wt%), and wrapped around the distal end of the MTR. 20 μl of a PVA solution (30 wt.%, Di H$_2$O) was ultimately applied to each corner of the fabric to secure it into position.

The potential clinical application of the resulting drug-loaded MTR was then illustrated by navigating it into the pancreatic duct of an anatomical ultrasoft phantom (Ecoflex GEL, Smooth-On), specifically the duodenum-pancreatic duct. The platform [54] consisted of a KUKA LBR iiwa 14 robot (KUKA, Germany), manipulating one external permanent magnet (EPM, cylindrical permanent magnet with a diameter and length of 101.6 mm and a magnetic moment of 970.1 $Am^2$ (Grade N52)). The MTR was deployed from the tool channel of a duodenoscope (JF-130, Olympus), then the EPM was positioned to apply a pulling magnetic field on the fabric-coated MTR, allowing it to travel into the pancreatic duct. PBS solution was added to the pancreatic duct. Upon positioning of the fabric-coated MTR, the US was applied directly to the phantom toward the pancreatic duct. US was applied for 3 min (pulsed mode, 50 % duty cycle, 100 Hz, with 1 MHz US frequency), after which time the MTR was removed, the PBS solution was extracted, and the 280 nm absorbance was measured. Silicones such as Ecoflex™ and Dragon Skin typically exhibit an acoustic impedance ranging between $1.03 - 1.07 \times 10^6$ N·s·m$^{-3}$.[92] This impedance is relatively low compared to soft tissues,[93] suggesting limited acoustic interaction between the MTR and the US.

### 5.10. Statistical analysis

All data is presented as the mean ± the standard deviation, with a sample size of 3 unless otherwise stated. Significant differences between data are determined using the two-tailed Student's t-test, where significance was defined as p-values ≤ 0.05. Statistical analysis was carried out using OriginPro Software.

**Supporting Information**

Supporting Information is available from the Wiley Online Library or from the author.

**Author Contributions**

Moorcroft planned and conducted experimentation, and wrote the manuscript with consultation from all authors.

Calmé conducted the magnetic tentacle robot experimentation, wrote the experimental protocol.

Brooker performed preliminary experimentation and offered consultation throughout the project.

Valdastri developed the magnetic robot used for demonstration of the magnetic tentacle robot.

Harris acquired funding, co-conceived the project and provided supervision.

Russell offered supervision and experimental consultation.

Tronci co-conceived and supervised the project, and assisted writing the manuscript.

**Acknowledgements**

The authors would like to thank the following agencies for financial support: EPSRC (Engineering and Physical Sciences Research Council) under grants EP/P027687/1 and EP/V009818/2; ERC (European Research Council) under grant number 818045; the University of Leeds; and the Clothworkers' Company. The authors would also like to thank the Leeds electron microscopy and spectroscopy centre (LEMAS), Mohammed Asaf, Emma Black, and Peter Lloyd, for their assistance with this project.

**Conflict of Interest Statement**

The authors declare that they have no conflict of interest.

**Data Availability Statement**

Data is available at: https://doi.org/10.5518/1655

**Ethical Statement**

No experimentation was performed with human or animal subjects.

# Supporting Information

**Ultrasound-triggered release of anticancer nanoparticles from electrospun fabrics integrated with soft robotic tentacles**

*Samuel C. T. Moorcroft,* *Benjamin Calmé, Charles Brooker, Pietro Valdastri, Russell Harris, Stephen J. Russell, Giuseppe Tronci*

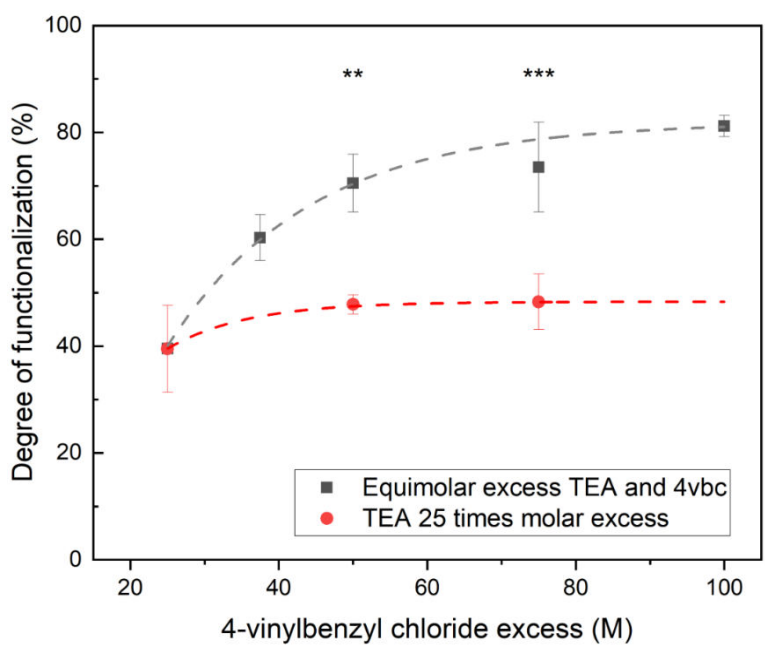

**Fig. S1.** The degree of functionalization of native porcine gelatin with 4-vinylbenzyl chloride (4vbc) when varying the molar excess of triethylamine (TEA) and 4vbc during the functionalization process. 4vbc excess is varied between 25 and 100 times with equimolar (grey) or 25 times (red) excess TEA. Error bars indicate the standard deviation (n = 3). p-values indicate: ** ≤ 0.01, *** ≤ 0.001.

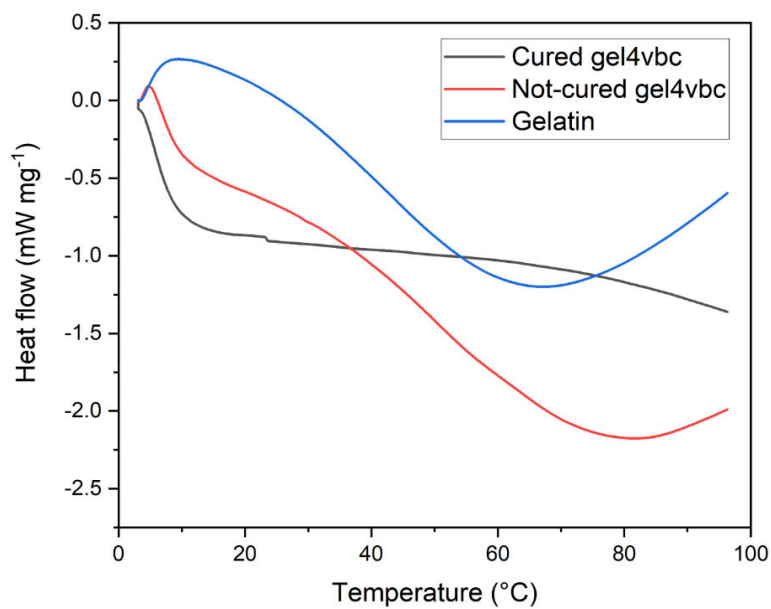

**Fig. S2.** Differential scanning calorimetry of native porcine gelatin (blue), gel4vbc (f = 40%, red), and c) gel4vbc (f = 40%) UV-cured for 30 min when hydrated in a 1 wt% Irgacure solution (black).

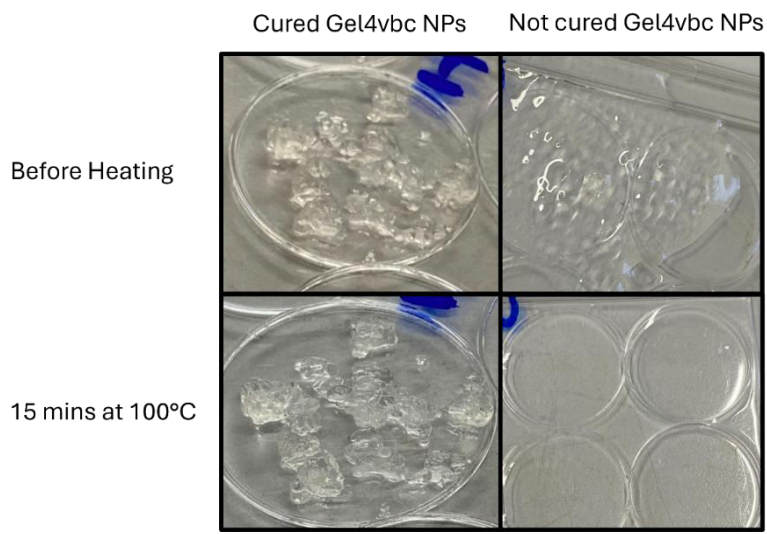

**Fig. S3.** Images of gel4vbc NPs with and without UV-curing. NPs were placed into an oven at 100°C for 15 mins.

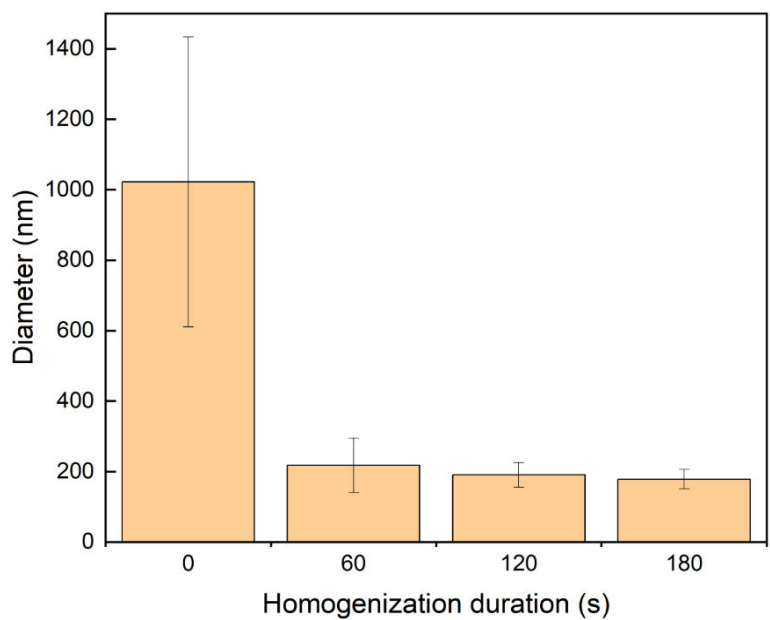

**Fig. S4.** NP size at varying homogenization durations at 8000 rpm, ascertained using multi-angle dynamic light scattering. NPs made using 10 wt% gel4vbc in aqueous phase at 1:32 gel4vbc solution: diethyl ether vol. ratio and 0.1 wt% SPAN80.

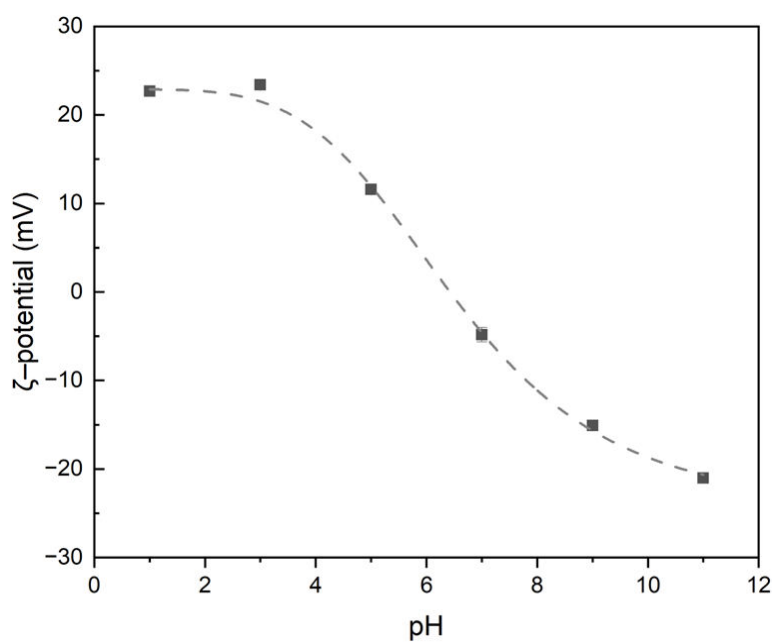

**Fig. S5.** Zeta potential measurements of gel4vbc NPs dispersed into the following buffers at 10mM: HCl-KCl (pH 1), citrate-phosphate (pH 3, 5), Tris (pH 7, 9), phosphate-NaOH (pH 11). Measurements were made in triplicate.

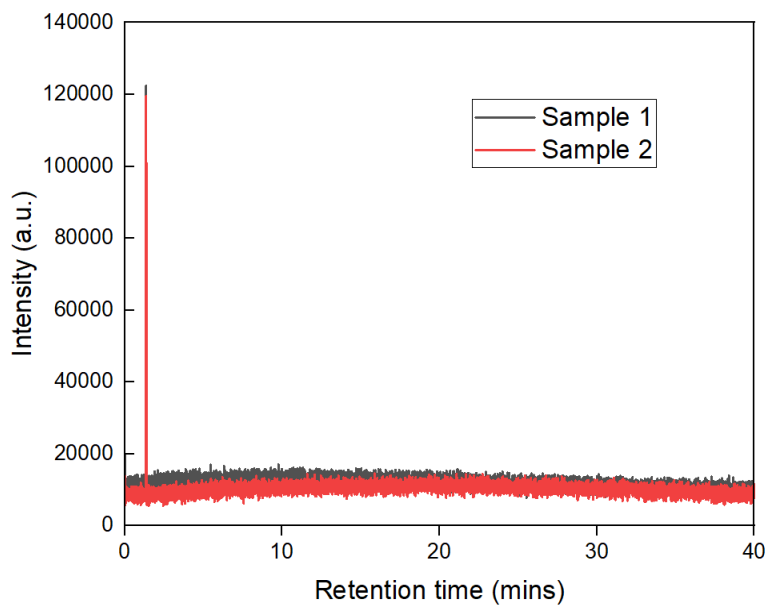

**Fig. S6.** Mass spectroscopy – gas chromatography traces of gel4vbc NPs in 10 mM HCl containing 0.1 wt% Irgacure. The sample was heated to 250°C and flowed through the machine using argon carrier gas, as seen through the peak at 1.36 min.

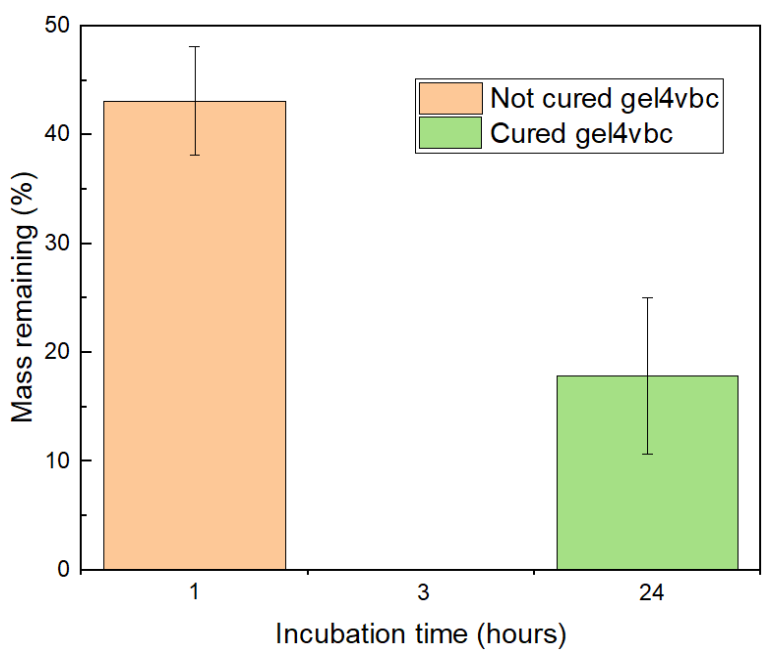

**Fig. S7.** Mass remaining of cured (orange) and non-cured gel4vbc (green) gels when incubated at 37°C in 1 CDU/ml Collagenase A. Error bars indicate standard deviation (n = 2).

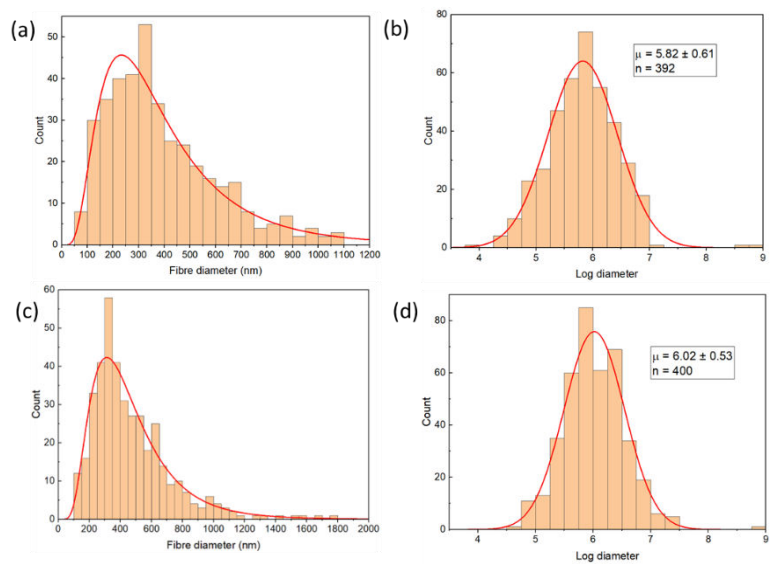

**Fig. S8.** a) Histogram plotting the diameter of PCL electrospun fibers. The red line acts as an eye-guide showing a positive data skew. b) Histogram plotting the base-10 logarithmic distribution of the PCL fiber size. The red line denotes a normal distribution. c) Histogram plotting the diameter of PCL electrospun fibers coated in PVA and NPs (0.5 and 1 wt%, respectively) and ultrasound irradiated. Ultrasound was applied for 3 min at 3 W cm$^{-2}$ at 1 MHz, in pulsed mode with a duty cycle of 50 % and at 100 Hz. The red line acts as an eye-guide showing a positive data skew. d) Histogram plotting the base-10 logarithmic distribution of the polymer-coated and US-treated PCL fiber size. The red line denotes a lognormal distribution. n = 392 for graphs a and b, n = 400 for graphs c and d.

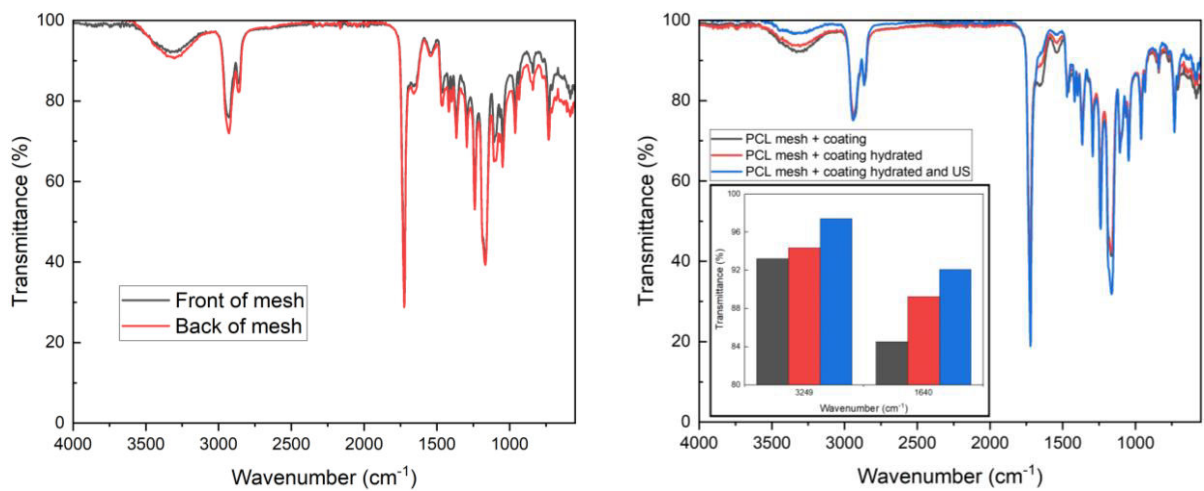

**Fig. S9.** Left) Fourier transform infrared spectra (FTIR) of PCL fabric with NP and PVA (1 and 0.5 wt%, respectively) coating from the front (grey) and back (red) of the fabric. Right) FTIR spectra of PCL fabric with PVA and NP coating (grey) and equivalent fabrices incubated in PBS at 37°C for 15 min without (red) and with subsequent ultrasound irradiation (blue). Inset: The transmittance values at the O-H band at 3249 cm$^{-1}$ and the C=O stretching at 1640 cm$^{-1}$. Ultrasound was applied for 3 min at 3 W cm$^{-2}$ at 1 MHz, in pulsed mode with a duty cycle of 50 % and at 100 Hz.

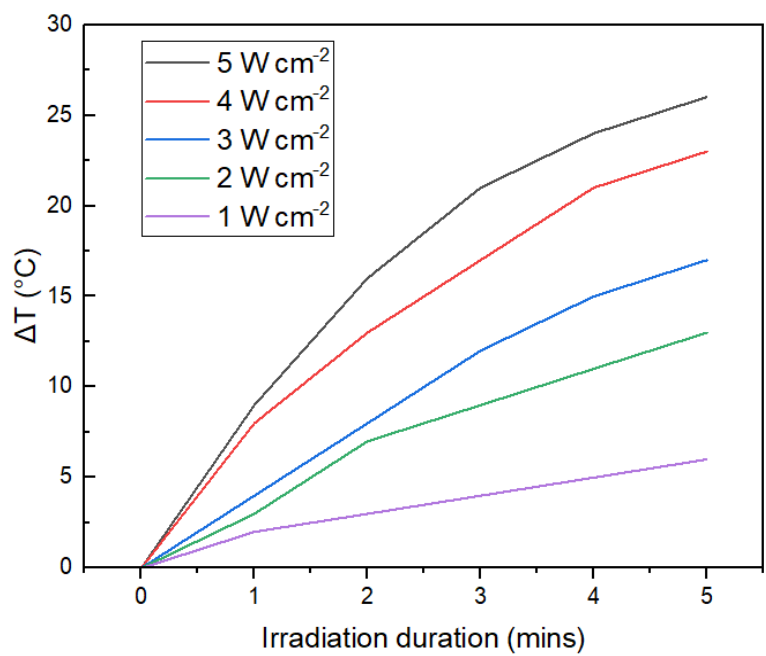

**Fig. S10.** Temperature increase from 22°C when 10 ml of PBS is irradiated with ultrasound from 1 – 5 W cm$^{-2}$.

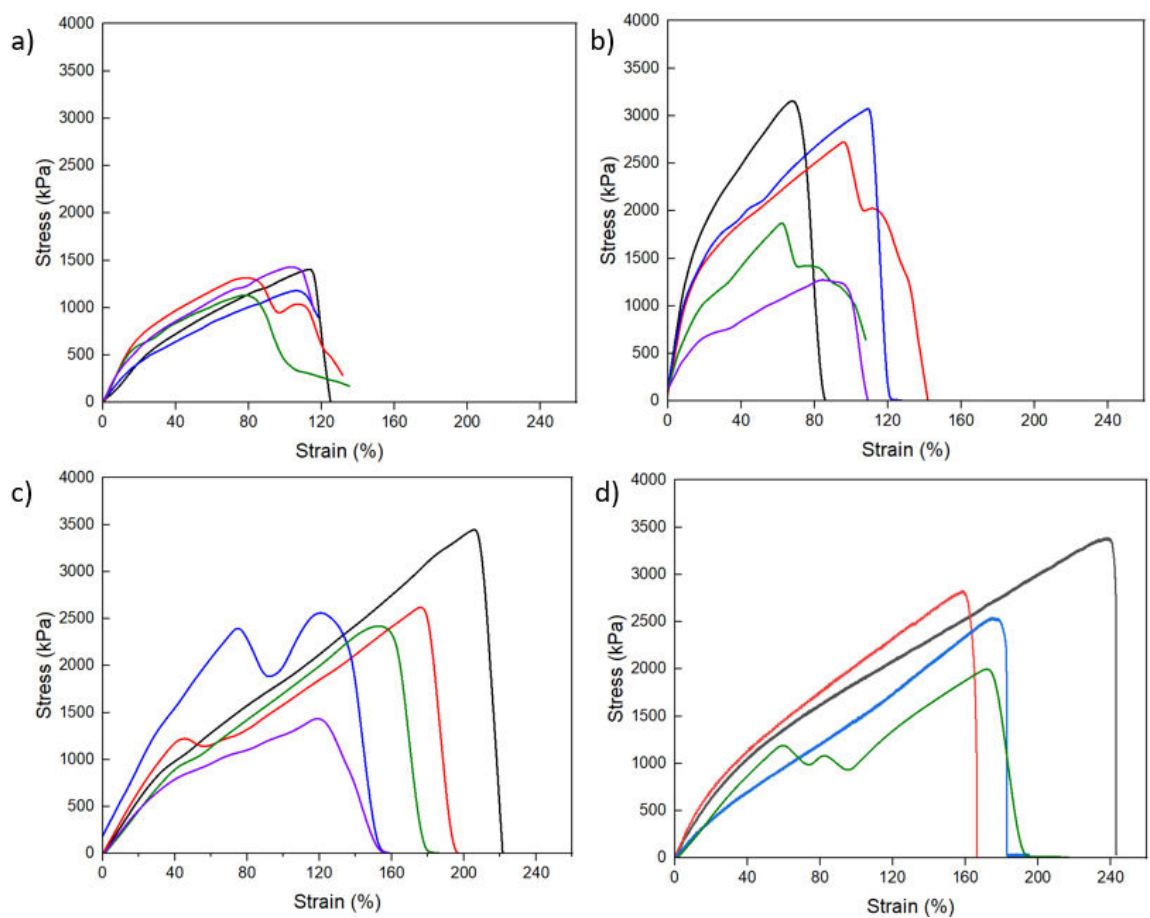

**Fig. S11.** The stress-strain graphs plotting the PCL fabric alone (a), PCL fabric with NP-PVA coating (b), and NP-PVA coated fabric when incubated in PBS for 1 hr without (c) and with (d) US. US was applied for 3 min at 3 W cm$^{-2}$ at 1 MHz, in pulsed mode with a duty cycle of 50% and at 100 Hz.

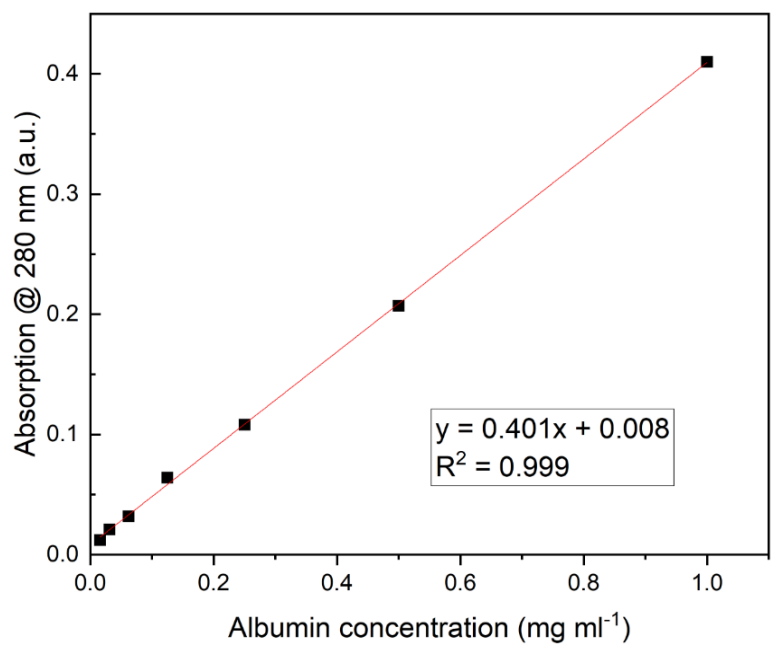

**Fig. S12.** Plot of 280 nm absorption against albumin concentration.